\input harvmac
\newcount\figno
\figno=0
\def\fig#1#2#3{
\par\begingroup\parindent=0pt\leftskip=1cm\rightskip=1cm
\parindent=0pt
\baselineskip=11pt
\global\advance\figno by 1
\midinsert
\epsfxsize=#3
\centerline{\epsfbox{#2}}
\vskip 12pt
{\bf Fig. \the\figno:} #1\par
\endinsert\endgroup\par
}
\def\figlabel#1{\xdef#1{\the\figno}}
\def\encadremath#1{\vbox{\hrule\hbox{\vrule\kern8pt\vbox{\kern8pt
\hbox{$\displaystyle #1$}\kern8pt}
\kern8pt\vrule}\hrule}}

\overfullrule=0pt

\Title{\vbox{\baselineskip12pt
\hbox{hep-th/0005007}
\hbox{TIFR-TH/00-21}
}}
{\vbox{\centerline{A Note on Supergravity Duals of}
\centerline{Noncommutative Yang-Mills Theory}}}
\smallskip
\centerline{Sumit R. Das~\foot{das@theory.tifr.res.in},
and Bahniman Ghosh~
\foot{bghosh@theory.tifr.res.in}}
\smallskip
\centerline{{\it Tata Institute of Fundamental Research}}
\centerline{\it Homi Bhabha Road, Mumbai 400 005, INDIA}
\bigskip

\medskip

\noindent
A class of supergravity backgrounds have been proposed as
dual descriptions of strong coupling large-N noncommutative Yang-Mills
(NCYM) theories in $3+1$ dimensions.  However calculations of
correlation functions in supergravity from an evaluation of relevant
classical actions appear ambiguous. We propose a resolution of this
ambiguity.  Assuming that {\it some} holographic description exists -
regardless of whether it is the NCYM theory - we argue that there
should be operators in the holographic boundary theory which create
normalized states of definite energy and momenta. An operator version
of the dual correspondence then provides a calculation of correlators
of these operators in terms of bulk Green's functions.  We show that
in the low energy limit the correlators reproduce expected answers of
the ordinary Yang-Mills theory.

\Date{April, 2000}

\def\tB{{\tilde B}}

\def\tx{{\tilde{x}}}
\def\hg{{\hat{g}}}
\def\vx{{\vec{x}}}
\def\cO{{\cal{O}}}
\def\vk{{\vec{k}}}
\def\log{{\rm{log}}}
\def\tB{{\tilde{B}}}
\def\hB{{\hat{B}}(ka)}
\def\ha{{\hat{a}}}
\def\cG{{\cal{G}}}

\newsec{Introduction}

Noncommutative gauge theories appear as limits of D-brane open
string theories in the presence of nonvanishing NSNS B fields
\ref\douglashull{A. Connes, M. Douglas and A. Schwarz,
JHEP 9802 (1998) 003, hep-th/9711162; M. Douglas and C. Hull, JHEP 02
(1998) 008, hep-th/9711165.},\ref\noncom{Y.K.E. Cheung and M. Krogh,
Nucl. Phys. B528 (1998) 185, hep-th/9803031; F. Ardalan, H. Arfaei and
M.M. Sheikh-Jabbari, JHEP 02:016 (1999), hep-th/9810072;
C.S. Chu and P.M. Ho,
Nucl. Phys. B550 (1999) 151, hep-th/9812219; 
M.M. Sheikh-Jabbari, Phys. Lett. B 455 (1999) 129, hep-th/9901080 and
JHEP 06 (1999) 015, hep-th/9903205; V. Schomerus, JHEP 9906:030
(1999), hep-th/9903205; D. Bigatti and L. Susskind,
hep-th/9908056.}. In \ref\seibergwitten{N. Seiberg and E. Witten, JHEP
9909:032 (1999), hep-th/9908142.} a precise decoupling limit
was defined in which the string tension is scaled to infinity and the 
closed string metric is scaled to zero keeping the dimensionful
NSNS B field fixed, and it was shown that in this limit the
open string theory on D-branes reduces to precisely noncommutative
Yang-Mills (NCYM) theory. Furthermore, NCYM appears naturally in the
IKKT matrix theory \ref\kawaietal{H. Aoki, N. Ishibashi, S. Iso,
H. Kawai, Y. Kitazawa and T. Tada, hep-th/9908141;   
N. Ishibashi, S. Iso, H. Kawai and
Y. Kitazawa, hep-th/9910004; I. Bars and D. Minic, hep-th/9910091; 
J. Ambjorn, Y. Makeenko, J. Nishimura and 
R. Szabo, hep-th/9911041; S. Iso, H. Kawai and Y. Kitazawa,
hep-th/0001027.}, a connection which has led to further insight.
Noncommutative field theories have novel perturbative behaviour
including an IR/UV mixing \ref\iruv{S. Minwalla, M. Raamsdonk and
N. Seiberg, hep-th/9912072; 
I. Ya. Aref'eva, D.M. Belov and A.S. Koshelev, hep-th/9912075
and hep-th/0001215;
M. Raamsdonk and N. Seiberg,
hep-th/0002186; M. Hayakawa, hep-th/9912094,hep-th/9912167;
W. Fischler, E. Gorbatov, A. Kashani-Poor, S. Paban, P. Pouliot and
J. Goomis, hep-th/0002067; A. Matusis, L. Susskind and N. Toumbas,
hep-th/0002075, O. Andreev and H. Dorn, hep-th/0003113;
Y. Kiem and S. Lee, hep-th/0003145; 
I. Ya. Aref'eva, D.M. Belov  A.S. Koshelev and O.A. Rytchkov,
hep-th/0003176;
A. Bilal, C. Chu and R. Russo,
hep-th/0003180; J. Gomis, M. Kleban, T. Mehen, M. Rangamani and
S. Shenker, hep-th/0003215; A. Rajaraman and M. Rozali,
hep-th/0003227; 
H. Liu and J. Michelson, hep-th/0004013;
G. Arcioni, J. Barbon, J. Goomis and
M. Vazquez-Mozo, hep-th/0004080; F. Zanora, hep-th/0004085; J. Gomis,
K. Landsteiner and E. Lopez, hep-th/0004115.}.

It is natural to ask whether the large-N limit of NCYM is dual to
closed string theory in some background, analogous to the duality of
usual Yang-Mills theory to string theories \ref\adscft{J. Maldacena,
Adv.Theor. Math. Phys. 2 (1998) 231, hep-th/9711200},\ref\corr{S. Gubser,
I. Klebanov and A. Polyakov, Phys. Lett. B428 (1998)
105,hep-th/9802109; E. Witten, Adv. Theor. Math. Phys. 2 (1998) 253,
hep-th/9802150; 
D. Freedman, S. Mathur, A. Matusis and
L. Rastelli, Nucl. Phys. B546 (1999) 96, hep-th/9804058.}. 
In fact, in \ref\hashimoto{A. Hashimoto and
N. Itzhaki, hep-th/9907166.} and \ref\maldarusso{J.  Maldacena and
J. Russo, hep-th/9908134.} supergravity duals of strong 't Hooft
coupling limits of NCYM were proposed. Further evidence for such
duality came from the study of D-instantons in such supergravity
backgrounds and their relationship to instantons of the NCYM
\ref\dkt{S.R. Das, S. Kalyana Rama and S. Trivedi,
JHEP 0003 : 004, 2000; hep-th/9911137.}. Aspects of such duality
have been explored in \ref\duals{M. Li, hep-th/9909085; A. Dhar, G. Mandal,
S. Wadia and K.P. Yogendran, hep-th/9910194; T. Harmark and
N. Obers, JHEP 0003:024 (2000), hep-th/9911169;
J. Liu and S. Roy,
hep-th/9912165; J. Barbon and A. Pasquinucci, hep-th/0002187;
N. Ishibashi, S. Iso, H. Kawai
and Y. Kitazawa, hep-th/0004038.}

However, the nature of this duality has been rather confusing.
In usual AdS/CFT duality, there is a well defined connection
between boundary correlators in supergravity and correlators
of local operators of the boundary conformal field theory. Consider
$AdS_{d+1}$ with a metric
\eqn\one{ ds^2 = {1\over z^2}(-dt^2 + dz^2 + d\vx \cdot d\vx)}
where $ \vx = x^i , i = 1 \cdots  (d-1) $. The boundary of this
space is at $z = z_0$ with $z_0 << 1$. In the strong 't Hooft coupling limit
the correspondence may be written as
\eqn\two{e^{-\Gamma_{eff}[\phi^i_0 (x)]}
= \int {\cal D}A_\mu~e^{-S_{CFT;z_0} - \sum_i
\int dx \phi^i_0 (x) {\cal O}_i (x)}}
where $S_{CFT;z_0}$ is the action of a $d$ dimensional theory
with a cutoff $z_0$. $\Gamma_{eff}[\phi^i_0 (x)]$ is the effective
action of supergravity in the background \one\ as a functional
of the values of the various fields $\phi^i (z,x)$ on the
boundary, 
\eqn\three{\phi^i (z_0,x) = z_0^{\Delta_i - d}\phi^i_0(x)}
 $ {\cal O}_i (x) $ is the operator in the boundary theory which
is dual to the field $\phi^i$.
This particular behaviour of the fields near the boundary given
in \three\ is guaranteed by the conformal isometries of the
$AdS$ space-time and $\Delta_i$ turns out to be the conformal
dimension of the operator ${\cal O}_i (x)$. Thus in this
case there is a holographic relationship between the bulk theory
and a {\it local} field theory on the boundary.

In \maldarusso\ a similar approach was tried for the spacetimes
proposed to be duals of NCYM. It was found that to make sense of the
results, the relationship between the boundary values of the
supergravity fields and the sources of some operators in the proposed
dual boundary theory is necessarily nonlocal in position space. In
other words the relationship between momentum space quantities involve
nontrivial functions of momentum.  Similar momentum dependent factors
were encountered in the study of one point functions in instanton
backgrounds in \dkt. It is not {\it a priori} clear what these factors
should be and the entire procedure appears ambiguous since we can 
get any answer we want by considering different renormalization factors.

In this paper we assume that a holographic description of the bulk
theory in such backgrounds exists in terms of a theory living on the
boundary - which may or may not be the NCYM theory.  This boundary
theory will be generically nonlocal and there would not be local
operators creating physical states.  However, because of translation
invariance, there are well defined operators which create {\it
normalized} states of definite energy and momenta.  A dual
correspondence in this context means that such states are to be
identified with states of supergravity modes propagating in this
background.  Consequently, {\it bulk} correlation functions in
supergravity, with points taken to lie on the boundary, define a set
of momentum space boundary correlators unambiguously. This gives
a supergravity prediction for such correlators which should reduce
to usual Yang-Mills correlators in the low energy limit. We show that
this is indeed true.

One of the motivations to study the dual correspondence in the
decoupling limit of \seibergwitten\ is that some of the dual space-times
are asymptotically flat in terms of the Einstein metric. The hope
is that some understanding of the holographic correspondence 
may teach us something about holographic relationships in other
asymptotically flat spaces. In fact there is a close relationship
between the type of problems encountered here and those encountered
in the holographic correspondence for NS five branes \ref\seibergmin{
S. Minwalla and N. Seiberg, JHEP 9906 (1999) 007, hep-th/9904142}.
Here again the dual spacetimes are asymptotically flat and
the boundary theory is nonlocal. Our prescription of reading off
correlators from bulk Green's function is in fact closely related to
the procedure adopted in \seibergmin\ where the correlators are
identified with relevant S-matrix elements.

In section 2 we present the supergravity solutions we are dealing with
and the equations for decoupled modes propagating in this background.
In section 3 we calculate the classical action for this
on-shell mode as in \maldarusso\ and state the ambiguity one encounters
in trying to extract boundary correlators from this action \foot{We correct
some errors in the treatment of \maldarusso.}. In section 4 we give the
operator version of the standard AdS/CFT correspondence for $B = 0$ and
show how boundary correlators of operators creating normalized
states in momentum space can be read off from bulk Green's function.
In section 5 we repeat this for our backgrounds with $B \neq 0$ and
show how boundary correlators of momentum space operators can be
obtained unambiguously. Section 6 contains concluding remarks.

\newsec{The supergravity solution and modes}

We will consider two kinds of IIB supergravity backgrounds with nonzero
$B$ fields obtained in \maldarusso. The D3 brane has a worldvolume
along $(x^0 \cdots x^3)$. 

The first kind of background has a nonzero NSNS $B_{23}$ with all
other components set to zero. In the decoupling limit which
corresponds to the low energy limit of \seibergwitten\ the string
metric is 
\eqn\aone{ds^2 = \alpha ' R^2[u^2 (- dx_0^2 + dx_1^2) + {u^2 \over 1+ a^4 u^4}
(dx_2^2 + dx_3^2) + {du^2 \over u^2} +d\Omega_5^2]}
where the dilaton $\phi$, NS field $B$ and the RR 2-form field
$\tB$ and the five form field strength $F$ are given by
\eqn\atwo{\eqalign{&e^{2\phi} = {g^2 \over 1+ a^4 u^4}~~~~~~~~~B_{23}
= {\alpha ' R^2 \over a^2}{a^4u^4 \over 1+ a^4 u^4}\cr
& \tB_{01} = {\alpha ' a^2 R^2 \over g}u^4~~~~~~~~F_{0123u}
= {\alpha '^2 \over g(1+ a^4u^4)}\partial_u (u^4 R^4)}}
where $R^4 = 4\pi gN$ and $g$ is the {\it open} string coupling.
In the infrared, $u \rightarrow 0$ the space time is 
$AdS_5 \times S^5$.

In this background, the graviton fluctuation $h_{01}$ with 
zero momenta along $x^0, x^1$ and zero angular momenta along the
$S^5$ satisfy a simple decoupled equation.
With $\phi = g^{00}h_{01}$ this is
\eqn\athree{ \partial_\mu(\sqrt g e^{-2\phi} g^{\mu\nu}\partial_\nu
~\phi) = 0}
which becomes in terms of modes
\eqn\afour{\phi (u,x_2,x_3) = \int [{d^2 k \over (2\pi)^4}]
\phi (\vk,u) e^{-ik_2x^2 - ik_3 x^3}}
\eqn\afive{\partial_u(u^5 \partial_u \phi (\vk,u))
- k^2 u(1 + a^4 u^4)\phi (\vk,u) = 0}
where in \afive\ $k^2 = k_2^2 + k_3^2$.
Such zero energy perturbations do not make sense in Lorentzian signature.
We will therefore work in the euclidean signature.

The second kind of background has self dual $B$ fields and has in addition
a nontrivial axion $\chi$. The decoupling limit 
solution in euclidean signature is
given by the Einstein metric $ds_E^2$ 
\eqn\eleven{ds_E^2 = {\alpha ' R^2 \over \sqrt{\hat g}}
[(f(u))^{-1/2}(d\tx_0^2 + \cdots + d\tx_3^2) +
(f(u))^{1/2} (du^2 + u^2 d\Omega_5^2)]}
while the other fields are
\eqn\elevena{\eqalign{
&e^{-\phi_0} =  {1\over \hg} - i~\chi_0 = {1\over \hg} u^4 f(u)\cr
& \tB_{01} = \tB_{23} = -{i\over \hg} B_{01} = -{i\over \hg} B_{23}
= - {i \alpha ' a^2 R^2 \over \hg}~(f(u))^{-1}\cr
&F_{0123u} = {4i(\alpha ')^2 R^4\over \hg u^5}(f(u))^{-2}}}
where
\eqn\twelve{f(u) = {1\over u^4} + a^4,~~~~~~~
R^4 = 4\pi \hg N}
An interesting feature of this solution is that the space-time is
asymptotically flat (in Einstein metric) 
even in the decoupling limit. In fact \eleven\ is
exactly the full D3 brane metric. Furthermore
near the boundary $u = \infty$ the string coupling vanishes.

In the background \eleven\ - \twelve, it was shown in
\dkt\ that there are special
fluctuations which satisfy decoupled equations. These are
fluctuations of dilaton $\delta \phi$ and axion $\delta \chi$
which obey the condition
\eqn\thirteen{\delta \chi + i e^{-\phi}\delta \phi =0.}
The corresponding equation for the fluctuation $\delta\phi$ is
\eqn\fourteen{\nabla _E^2 (e^{\phi_0}\delta\phi) = 0}
where $\nabla _E^2$ is the laplacian in the Einstein
metric $ds_E^2$ given in \eleven. 
These fluctuations are now allowed to carry momenta along all
the brane worldvolume directions, but for simplicity we will
consider zero $S^5$ angular momenta.
The modes are
\eqn\fifteen{e^{\phi_0}\delta\phi (u,x) = \int
[{d^4 k \over (2\pi)^4}] \Phi (\vk,u)~e^{-i\vk \cdot \vx}}
where
\eqn\sixteen{\vk \cdot \vx = \sum_{i=0}^3 k_i x^i}
and the equation \fourteen\ once again becomes
\eqn\seventeen{\partial_u(u^5 \partial_u \Phi (\vk,u))
- k^2 u(1 + a^4 u^4)\Phi (\vk,u) = 0}
which is identical to \afive. In \seventeen, 
\eqn\eighteen{k^2 = \sum_{i=0}^3 (k_i)^2}

\newsec{Solutions and the boundary action}

The solutions of the equation \seventeen\ or \afive\ may be
written in terms of Mathieu functions \ref\gubhashi{S. Gubser and
A. Hashimoto, Comm. Math. Phys. 203 (1999) 325, hep-th/9803023.}.
Introducing the coordinates
\eqn\bone{ u = {1\over a} e^{-w}}
the two independent solutions may be chosen to be
\eqn\btwo{{1\over u^2} H^{(1)}(\nu,w+{i\pi\over 2})~,~
{1\over u^2} H^{(2)}(\nu,-w-{i\pi\over 2})}
Here the parameter $\nu$ is determined in terms of the combination
$(ka)$. It has a power series expansion given by
\eqn\bthree{ \nu = 2 - {i{\sqrt{5}}\over 3}({ka \over 2})^4
+ {7i \over 108 {\sqrt{5}}}({ka \over 2})^8 + \cdots} 
The Mathieu functions $H^{(i)}$ have the asymptotic property
\eqn\bfour{H^{(i)}(\nu,z) \rightarrow H^{(i)}_\nu(ka e^z)~~~~~z \rightarrow
 \infty}
where $H^{(i)}_\nu(z)$ denotes Hankel functions. 
Furthermore, in this region, only $(ka)<<1$ contribute significantly,
so that $\nu \sim 2$.
Thus near $u=0$
where the geometry is $AdS_5 \times S^5$ the solutions become the
standard solutions of the massless wave equation in $AdS$, viz
$(1/u^2)K_2 (k/u)$ and $(1/u^2)I_2 (k/u)$

>From the asymptotic
behavior of the Hankel functions it is clear that the solution
$H^{(1)}(\nu,w+{i\pi\over 2})$ is well behaved and goes to zero in the
interior of the spacetime at $u = 0$, while the solution 
$H^{(2)}(\nu,-w-{i\pi\over 2})$ is well behaved at $u= \infty$.
\eqn\bfive{\eqalign{&H^{(1)}(\nu,w+{i\pi\over 2})
\rightarrow e^{-i{\pi\over 2}(\nu+1)}{\sqrt{2\over \pi k a e^w}}
e^{-ka e^w} ~~~~~~w \rightarrow \infty ( u \rightarrow 0)\cr
&H^{(2)}(\nu,-w-{i\pi\over 2})
\rightarrow e^{i{\pi\over 2}(\nu+1)}{\sqrt{2\over \pi k a e^{-w}}}
e^{-ka e^{-w}}~~~~~~w \rightarrow -\infty (u \rightarrow \infty)}}

\subsec{Supergravity actions and correlators in the usual AdS/CFT 
correspondence}

Let us recall the standard way of obtaining correlators in the
AdS/CFT correspondence \corr\ using \two.
For a minimally coupled scalar field of mass $m$ in $AdS_{d+1}$
with a metric given by \one\ one considers a solution (in euclidean signature)
which is smooth in the interior
\eqn\fourteen{\phi(x,z) = \int [{d^dk \over (2\pi)^d}]
k^\nu~ z^{d/2}~K_\nu(kz) e^{-ik\cdot x}~\phi_0 (k)}
Here $x$ denotes all the four
directions $x = (\vx,t)$ and $k^2 = k_0^2 + \vk^2$.
One then computes the supergravity action after putting in a boundary
at $z = z_0 << 1$. Note that the fourier modes $\phi_0 (k)$ have been
defined so that as one approaches the boundary
\eqn\sixteen{{\rm Lim}_{z \rightarrow 0}\phi (z,k)
= z^{{d\over 2} - \nu} \phi_0 (k)}
$\phi_0(k)$ are then taken to be sources for operators $ O(k)$
conjugate to the supergravity field in the Yang-Mills theory living
on the boundary.
The action is purely a 
boundary term with the leading nonanalytic piece
\eqn\fifteen{S \sim \int [{d^dk \over (2\pi)^d}]
\phi_0(k)\phi_0(-k) k^{2\nu} \log (kz_0)}
The correspondence \two\ then leads to a two point function
of $\cO(k)$ with a leading nonanalytic piece
\eqn\fifteena{<{\cal O}(k) {\cal O}(-k)> \sim k^{2\nu}~\log (kz_0)}
which shows that the dimension of this operator is
\eqn\fifteenb{ \Delta = d/2 + \nu}
This specific relation between the sources and the boundary values
of the field is simple -
the power of the infrared cutoff which appears in \sixteen\
is $(d - \Delta)$.
By the IR/UV correspondence this is precisely the power of
{\it ultraviolet} cutoff necessary to add a perturbation
$\cO(k)$ to the boundary action.

\subsec{Supergravity actions for $B \neq 0$}

The proposal of \maldarusso\ is to consider $u = \infty$ as the
boundary also for the classical solutions in the presence of a
$B$ field. 
The solution to be used in calculating the classical action has
to be regular in the interior which means we have to take
\eqn\bsix{\phi_k (u) = {1\over u^2} H^{(1)}(\nu,w+{i\pi\over 2})
e^{i{\pi\over 2}(\nu+1)}~\phi_0 (k)}
Once again the classical action is a boundary term.
To evaluate this term we need to find the behavior of solution
\bsix\ for large values of $u$. This can be done using the relation
\eqn\bseven{H^{(1)}(\nu,w) = {1\over C(ka)}
[(\chi (ka) -{1\over \chi (ka)})H^{(1)}(\nu, -w)+
(\chi (ka) -{1\over \eta (ka)^2 \chi (ka)})H^{(2)}(\nu, -w)]}
where we have defined
\eqn\beight{ \eta (ka) = e^{i\pi\nu}~~~~~C(ka) = \eta (ka) - 
{1\over \eta (ka)}}
and the function $\chi (ka)$ has been defined in \gubhashi\ in terms
of relations between various Mathieu functions. Defining further the
functions
\eqn\bnine{A(ka) = \chi (ka) - {1\over \chi (ka)}~~~~~~B(ka) = \eta (ka)
\chi (ka)
- {1\over \eta (ka)\chi (ka)}}
the asymptotic form of the solution $\phi_k(u)$ becomes
\eqn\bten{\phi_k (u) \rightarrow
{1\over u^2 ~C(ka)}{\sqrt{2 \over \pi ka^2 u}}[iA(ka)~e^{ka^2u}
- \hB e^{-ka^2 u}] \phi_0 (k)~~~~~~(u \rightarrow \infty)}
where $\hB$ denotes the real part of $B(ka)$
\foot{The real part has to be taken in this asymptotic expansion
since the wave functions are real. This is similar to what happens in
the asymptotic
expansions of modified Bessel's functions.}. We will see shortly that
$A(ka)$ is purely imaginary so that the expression in \bten\ is real.
As expected the solution diverges exponentially at the boundary $u = \Lambda$
with $\Lambda >> 1$. The contribution to the classical action from this
solution, which becomes the boundary
term
\eqn\beleven{ [u^5 \phi_{-k} (u) \partial_u \phi_k (u)]_{u = \Lambda}}
is clearly divergent. Subtracting this infinite piece we get a term
where the exponentials cancel leaving a contribution
\eqn\bthirteen{ S_B = {5 \over 2 \Lambda}{2 \over \pi k a^2}{ i A(ka)~\hB
\over C^2 (ka)}~\phi_0 (k) \phi_0 (-k)}
Note that the 
boundary action given in \maldarusso\ has an error and differs from
the above by essentially a factor of $1/\Lambda$. 

Mimicking the standard procedure in the AdS/CFT correspondence we
might want to relate the functions $\phi_0 (k)$ to source terms in the
dual theory living on the boundary, so that derivatives of the
classical action with respect to these would give correlation
functions of the dual operators. So far, however, we have no clue
about this precise relationship.  The only guide we have at this stage
is that in the low energy limit $ka << 1$ the correlators should
reproduce the known answers in the $AdS_5 \times S^5$ case - for these
minimally coupled scalars the two point function should go as $k^4
\log (k)$.

\subsec{Low energy limits}

We need to find the low $ka$ expansion of the various functions. This
may be done using the results of \gubhashi\ as follows.
First note that the function $\eta (ka)$ is purely real, as follows
from the expression for $\nu$ in \bthree. The functions $A(ka), B(ka)$
and $C(ka)$ satisfy a unitarity relation \gubhashi\
\eqn\bfourteen{ |B(ka)|^2 = |A(ka)|^2 + |C(ka)|^2}
Using the reality of $\eta (ka)$ and hence $C$ it may be easily shown from
\bfourteen\ that $\chi (ka)$ must be a pure phase, so that $A(ka)$
is purely imaginary. Denoting
\eqn\bfifteen{ \eta (ka) = e^{\beta (ka)}~~~~~~~~\chi (ka) = e^{i\gamma (ka)}}
and using \bfourteen\ the various functions may be expressed in terms
of the function
\eqn\bsixteen{ P (ka) = {|C (ka)|^2 \over |B (ka)|^2}}
which is the absorption probability in the full D3 brane background
computed in \gubhashi. We give below some expressions which we
will need
\eqn\bseventeen{\eqalign{& {\hB \over iA (ka)} 
= [{P (ka) \cosh^2\beta (ka) - \sinh^2\beta (ka) \over 1 - P (ka)}]^{1/2}\cr
& {iA (ka) \over C (ka)} = [{1\over P (ka)} - 1]^{1/2}}}
The expansion of $P(ka)$ given in \gubhashi\ is
\eqn\beighteen{ P(ka) = A_0 (ka)^8 [1 + A_1 (ka)^4 + A_2 (ka)^4
\log (ka) + \cdots]}
where $A_i$ are numerical constants. The expansion of $\beta (ka)$ is of
the form
\eqn\bnineteen{ \beta (ka) = \beta_0 (ka)^4 [ 1 + \beta_1 (ka)^4
+ \beta_2 (ka)^8 + \cdots]}
where $\beta_i$ are numerical coefficients.
This leads to the following expansions
\eqn\btwenty{\eqalign{& {\hB \over iA (ka)}
\sim (ka)^4 [ 1 + \alpha_1 (ka)^4 + \alpha_2 (ka)^4 \log (ka)
+ \cdots]\cr
& {iA \over C (ka) } \sim {1\over (ka)^4}[ 1 + \gamma_1 (ka)^4
+ \gamma_2 (ka)^4 \log (ka) + \cdots]}}
where the coefficients $\alpha_i$ and $\gamma_i$ may be obtained
from the expansions in \beighteen\ and \bnineteen. Using these, the action 
$S_B$ is seen to behave as
\eqn\btwoone{ S_B \sim {1 \over a(ka)^5}[1 + (ka)^4 + (ka)^4 \log (ka)]
{1\over \Lambda} \phi_0 (k) \phi_0 (-k)}
for small momenta. Therefore if we define a renormalized boundary
value of the field
\eqn\btwotwo{ \Phi_0 (k) = F(ka) \phi_0 (k)}
such that
\eqn\btwothree{ F(ka) \sim {1\over (ka)^{5/2}}{1\over \Lambda^{1/2}}}
for small momenta, and declare that $\Phi_0 (k)$ are the sources 
which couple to the boundary theory operator dual to the bulk field
$\phi$, we would certainly get the correct low momentum behavior for the
nonanalytic piece of the two point function \foot{ In \maldarusso\
it is claimed that the two point function of the operators defined
there (which differs from ours) has the correct low energy behavior.
We havent been able to see how this follows}.

Clearly, this is an arbitrary procedure. Once we use momentum dependent
factors to renormalize fields, we can get any answer we want ! At this
stage there is no obvious principle which determines this factor. Any
statement about holography would be an empty statement.

\newsec{Bulk and Boundary Green's functions in $AdS/CFT$}

The usual $AdS_{d+1}/CFT_d$ correspondence may be understood in
terms of the modes of bulk field operators. Our treatment follows
\ref\banks{V. Balasubramanian, P. Kraus and  A. Lawrence,
Phys.Rev. D59 (1999) 04600, hep-th/9805171; 
T. Banks, M. Douglas, G. Horowitz and E. Martinec,
hep-th/9808016; V. Balasubramanian, P. Kraus, A. Lawrence and  S. Trivedi
,Phys.Rev. D59 (1999) 104021, hep-th/9808017} with some differences.
Consider quantization of a massive scalar field $\phi(z,\vx,t)$ 
of mass $m$ which is
minimally coupled to the AdS metric \one. The field has the following
mode expansion
\eqn\four{\phi(z,\vx,t) = {z^{d/2}\over 2 R^{(d-1)/2}}
\int_0^\infty d\alpha \int {d^{d-1}k \over (2\pi)^d}~
({\alpha \over \omega})^{1/2}~J_\nu (\alpha z)~[a(\vk,\alpha) 
e^{-i(\omega t - \vk \cdot \vx)} + (h.c.)]}
where 
\eqn\five{\omega^2 = \vk^2 +\alpha^2~~~~~
\nu = {1\over 2}(d^2 + 4m^2)^{1\over 2}}
The modes are normalized so that the annihilation/creation operators
satisfy the standard commutators
\eqn\six{ [a(\vk,\alpha),a^\dagger(\vk ',\alpha ')]
= \delta^{(d-1)}(\vk - \vk ')~\delta (\alpha - \alpha ')}
We can now make a change of integration variables from $(\alpha,\vk)$
to $(\omega,\vk)$ and define new operators \foot{There is a subtely here.
In Lorentzian signature we must have $\omega^2 > k^2$ so that the
range of integration over the four momenta is strictly restricted.
However, this fact has no consequence for two point functions. We
thank E. Martinec for discussions about this point.}
\eqn\six{b(\vk,\omega) = ({\omega \over \alpha})^{1/2}~a(\vk,\alpha)}
which satisfy the commutation relations
\eqn\sixa{[b(\vk,\omega),b^\dagger(\vk ',\omega ')]
= \delta^{(d-1)}(\vk - \vk ')~\delta (\omega - \omega ')}
and rewrite the expansion \four\ as
\eqn\foura{\phi(z,\vx,t) = {z^{d/2}\over 2 R^{(d-1)/2}}
\int d\omega \int {d^{d-1}k \over (2\pi)^d}~
J_\nu (\alpha z)~[b(\vk,\omega) 
e^{-i(\omega t - \vk \cdot \vx)} + (h.c.)]} 
In \foura\ $\alpha$ is determined in terms of $\omega$ and $\vk$ by the
relation \five. The states created by $b^\dagger(\omega,\vk)$,
denoted as $|\omega,\vk> = b^\dagger(\omega,\vk)|0> $ are
normalized according to $d$-dimensional delta functions, as follows
from \sixa,
\eqn\seven{<\omega,\vk | \omega ',\vk '> = 
\delta^{(d-1)}(\vk - \vk ')~\delta (\omega - \omega ')}

The holographic correspondence then implies that these states are
also states in the $d$ dimensional boundary theory and there are
composite operators which create these states. We can define these
operators in momentum space as
\eqn\eight{\cO (\omega,\vk) =
{2\pi \over R^{(d-1)/2}}(\omega^2-\vk^2)^{\nu/2}[
\theta (\omega) b(\omega,k) + \theta (-\omega) b^\dagger (-\omega,-\vk)]}
The overall power of $\alpha = (\omega^2-\vk^2)^{1/2}$ follows
from the fact that as we approach the boundary $z \rightarrow 0$
the radial wave function $J_\nu (\alpha z) \rightarrow (\alpha z)^\nu$.
This allows us to define in an unambiguous way a momentum space
correlation function of the boundary theory in terms of the boundary
values of the fourier transform of the Feynman Green's function of the
bulk field
\eqn\nine{<\cO (\omega,\vk)\cO(-\omega, -\vk)>
= {\rm Lim}_{z,z' \rightarrow 0}({\alpha^{2\nu}
\over \psi_{\omega,\vk}(z) \psi_{-\omega, - \vk}(z')})
G_F (z,z';\omega,k)}
where we have used translation invariance along the $\vx$ directions
and $\psi_{\omega, \vk}(z)$ denote the bulk radial wavefunctions 
\eqn\ten{\psi_{\omega,\vk}(z) = z^{d/2} J_\nu (\alpha z)}
The corresponding wavefunctions $\psi_{\omega,\vk}(t,\vx,z)
= z^{d/2} J_\nu (\alpha z)~e^{-i(\omega t - \vk \cdot \vx)}$
are normalized in terms of the standard Klein-Gordon norm
\eqn\tena{(\psi_{\omega,\vk}(\vx,t,z),\psi_{\omega ',\vk '}(\vx,t,z))
= \delta(\omega - \omega ') \delta (\vk - \vk ')}
We can now consider a Wick rotation to obtain a relation between
Euclidean Green's functions
\eqn\ninea{<\cO (k)\cO(-k)>_E
= {\rm Lim}_{z,z' \rightarrow 0}({k^{2\nu}
\over \psi_{k}(z) \psi_{-k}(z')})
G_E (z,z';\omega,k)}
where the wave functions are also rotated to euclidean space
and $k$ without a vector sign denotes the $d$ dimensional
euclidean momenta.

It may be easily checked that the euclidean bulk Green's function
leads to the euclidean correlators on the boundary obtained
according to the procedure of \corr. The euclidean bulk
Green's function is
\eqn\twelve{G_E(z,k;z',-k) =
(zz')^{d/2} K_\nu (kz) I_\nu (kz')~~~~~~~~z' < z}
Using the asymptotic form of the modified Bessel functions
for $z,z' \rightarrow 0$, we easily see that the leading
nonanalytic piece of the Green's function is
\eqn\thirteen{G_E (z,z',k) \rightarrow (zz')^{\nu + d/2}
~k^{2\nu}~{\rm log}~(kz)}
Using \ninea\ we get the boundary correlator
\eqn\thirteena{<\cO(k)\cO(-k)> \sim k^{2\nu} \log (kz)}
in agreement with \fifteena.

We can in fact define {\it local} operators on the
boundary by taking the fourier transform of the momentum space
field $\cO (\omega,\vk)$. In fact the power of $\alpha$ in
\eight\ has been chosen such that this boundary field is in fact
the boundary value of the bulk field upto the value of the bulk
wavefunction at the boundary
\eqn\eleven{{\rm Lim}_{z \rightarrow 0} \phi (z,\vx,t)
= (z)^{\nu + d/2} \cO (\vx, t)}
thus defining $\cO (\vx, t)$. The reason why this can be done
in an unambiguous fashion is that the wavefunctions decay as
the {\it same} power of $z$ regardless of the value of the
momenta. This in turn is a consequence of the conformal isometries
of $AdS$ space. For IIB supergravity on $AdS_5 \times S^5$ it
is known that the boundary operators thus defined are in fact
{\it local} gauge invariant operators of $N=4$ SYM theory.

\newsec{Bulk and Boundary Green's functions for $B \neq 0$}

For our supergravity backgrounds with $B \neq 0$ the
conjectured dual theory - NCYM - is not a local quantum field theory
in the conventional sense. The backgrounds we are dealing with are not 
asymptotically AdS - in fact
the second class of euclidean backgrounds are asymptotically flat. 
However we do have translation invariance
along the brane directions - so that physical states of the NCYM can be
still labelled by the energy and momenta. If this duality is indeed
correct, we should be able to represent such states by on shell states
of supergravity with the same values of energy and momentum - essentially
repeating the treatment of section 2. In this section we argue that
such a correspondence between operators in momentum space leads to
an unambiguous supergravity prediction of two point functions for
the corresponding operators in the NCYM theory. We will then verify that
this has the correct low energy behavior.

For the kind of fields which satisfy the minimally coupled massless
Klein Gordon equation as we have been studying the mode expansion in
our background may be written down in analogy to (2.6) with the Bessel
function being replaced by the appropriate Mathieu function. With
Lorentzian signature there are two independent wavefunctions which are
(delta function) normalizable, corresponding to incoming and outgoing waves
in the full D3 brane geometry. In the following we will consider only the
wave function whose euclidean continuation does not blow up at infinity -
this is the wave function given by
\eqn\jzero{\Psi_k (u) =  N(ka)~e^{-i{\pi \over 2}(\nu+1)} 
~{1\over u^2}H^{(2)}(\nu,-w)}
where $N(ka)$ is a normalization factor.
 The mode
expansion may be then written as
\eqn\jone{\phi(\vx,u,t) = \int {d^3 k \over (2\pi)^3}
{d\omega \over 2\pi}~N(ka)~{1\over u^2}H^{(2)}(\nu,-w)~e^{-{i\pi\nu\over2}}~
[e^{-i(\omega t - \vk \cdot \vx)}~\ha (\omega,k) + c.c.]}
$N(ka)$ is 
determined by requiring
that the operators $\ha (\omega, k)$ satisfy the standard commutation
relation
\eqn\jtwo{[\ha (\omega,k), \ha^\dagger (\omega ', k')]
= \delta^{(3)}(\vk - \vk ')\delta (\omega - \omega ')}
Of course there is no ambiguity in this expansion. Following the logic
of section 2 we then conclude that there are operators which create well
defined states and the two point function of these operators are related
to the bulk Green's functions by a relation similar to \ninea\
\eqn\jthree{<\cO (k)\cO(-k)>_E
= {\rm Lim}_{u,u' \rightarrow \infty}({k^4
\over \Psi_{k}(u) \Psi_{-k}(u')})
\cG_E (u,u';k)}
where $\cG_E$ is the euclidean Green's function in this background. 
(In this expression $k$ stands for the four vector)

The Green's function $\cG_E$ can be easily computed since we know the
two independent classical solutions - in fact this has already been
computed in \dkt\ to obtain the D-instanton solution in these
backgrounds. This is given by
\eqn\jfour{\eqalign{&\cG_E(u,u';k) = {\pi C(ka) \over 4i (uu')^2 A(ka)}
H^{(1)}(\nu,w+{i\pi\over 2})H^{(2)}(\nu, -w'-{i\pi\over 2})
~~~~u' > u \cr
&\cG_E(u,u';k) = {\pi C(ka) \over 4i (uu')^2 A(ka)}
H^{(1)}(\nu,w'+{i\pi\over 2})H^{(2)}(\nu, -w-{i\pi\over 2})
~~~~u > u'}}
Using the asymptotic expressions for the Mathieu functions (equation
\bten) one gets for $u' > u >> 1$
\eqn\jfive{\cG_E(u,u';k) = \cG_0 (u,u';k) +
{1\over 2ka^2}{1\over (uu')^{5/2}}~{i\hB \over A(ka)}~
e^{-ka^2 (u + u')}}
where $\cG_0$ denotes the Green's function in {\it flat} space
\eqn\jsix{\cG_0 (u,u';k) = {1\over 2ka^2}
{1\over (uu')^{5/2}}
e^{-ka^2 (u' - u)}}
In \jthree\ we will have to substitute this asymptotic $\cG_E$. In view of the
fact that the Green's function became a sum of the free piece and a
``connected'' piece suggests that it is natural to subtract out the
free piece in \jthree. We will adopt this prescription.

Note that the expression \jthree\ is unambiguous - there is no room
for momentum dependent wave function renormalizations here. The Green's
function $\cG_E$ is completely determined once the wave equation is
known, including all normalizations and the wave functions are determined
upto phases which in any case cancel in this expression.

If \jthree\ is to make any sense, the $u$ dependence should cancel in
the right hand side. From the asymptotic form of $H^{(2)}$ given in
\bfive\ and the form of the connected bulk Green's function in \jfive\
we see that this indeed happens. The final answer is
\eqn\jseven{<\cO(k)\cO(-k)> = k^4 {i \hB  \over A(ka)}~
({1\over N(ka)})^2}

We do not know how to obtain an explicit expression for $N(ka)$ for
all $ka$ - though we emphasize that it can be obtained in principle.
However it is straightforward to find the behavior of this normalization
factor for small values of $ka$. This is because in this regime the
various Mathieu functions become Bessel functions. Using 
the relation (notations are those of \gubhashi)
\eqn\jeight{H^{(2)}(\nu,-z) = {\eta (ka) \over \chi (ka)}H^{(1)}(\nu,z)
-{2 A(ka) \over C(ka)} J(\nu,z)}
we see that in this regime
\eqn\jnine{H^{(2)}(\nu,-w-{i\pi\over 2}) \rightarrow
{2 iA(ka)\over C(ka)} e^{i{\pi\over 2}(\nu+1)}~I_\nu (k/u)}
Since $I_\nu$ is the euclidean continuation of the normalizable solution
in the $AdS$ limit we now know that for $ka << 1$ we must have
\eqn\jten{ N(ka) \rightarrow {C(ka) \over iA(ka)}}
Plugging in the small $ka$ expansions for the various functions,
given in \btwenty\ we easily verify that the leading nonanalytic
term for small $ka$ is given by
\eqn\jeleven{< \cO(k) \cO(-k)> \sim k^4 \log (ka)}
which is the answer in the absence of a $B$ field.

The fact that we have obtained the correct low $(ka)$ behavior is
nontrivial. This is because we have taken the boundary at $u = \infty$
{\it before taking any low energy limit}. We consider this result to
be an evidence in favor of the holographic correspondence proposed 
in \maldarusso.

The procedure adopted above to obtain the correlation functions is
in fact similar to the procedure adopted in \seibergmin\ for the case
of NS five branes where again one has an asymptotically flat geometry
in the decoupling limit. In this work the two point function is related
to the S-matrix. This can be of course obtained from the euclidean
Green's function using a reduction procedure. Our procedure is of
course valid for the standard AdS case where the space-time is not
asymptotically flat.

\newsec{Conclusions}

The boundary theory proposed to be dual to the supergravity
backgrounds we have considered in this paper - viz. NCYM theory - is
nonlocal.  It is likely that there are no local gauge invariant
operators (in terms of the usual noncommutative gauge fields ) 
in this theory and correlation functions in position space
do not make any obvious sense. In this paper we have argued that
whether or not the boundary theory is indeed NCYM theory, states are
still specified by values of the energy and (three dimensional)
momenta. Thus there must be {\it some} operators $\cO(k)$ in momentum
space which create such states normalized in the standard fashion.
The holographic correspondence identifies these states with normalized
states in supergravity. We have shown that this implies an unambiguous
prediction of the correlators of $\cO(k)$. We verified that for small
momenta they reproduce the expected result for usual YM theory.

If the boundary theory is indeed the NCYM theory, it is crucial to
find the momentum space dual operators.  In the standard $AdS/CFT$
correspondence one useful way to read off the dual operators is to
consider the coupling of a three brane to background supergravity
fields, using, e.g. a nonabelian version of the DBI-WZ action
\ref\dastriv{S.R. Das and S. Trivedi, Phys. Lett. B445 (1998) 142,
hep-th/9804149; S. Ferrara, M. Lledo and A. Zaffaroni, Phys. Rev. D58
(1998) 105029, hep-th/9805082.}.  One possibility is to explore these
couplings in the presence of a $B$ field and in the low energy limit
of \seibergwitten. Rewriting DBI-WZ actions will not be sufficient in
this case since the nontrivial modifications come when the
supergravity backgrounds carry momenta in the brane direction so that
derivatives of gauge fields are important.  Some couplings of closed
string modes to open string modes in the presence of $B$ field have
been studied in \ref\hyunkiemlee{S. Hyun, Y. Kiem, S. Lee and C.Y. Lee,
hep-th/9909059}. However it remains to be seen whether one could
extract the couplings in terms of the standard fields of the
NCYM theory. In particular it appears that the effect of $B$ fields
may not be all encoded in the star product once closed string couplings
are included.

\newsec{Acknowledgements} We would like to thank 
R. Gopakumar, Y. Kiem, J. Maldacena, S. Minwalla and S. Trivedi for
discussions.  S.R.D. would like to thank the String Theory Group at
Harvard University for hospitality during the final stages of this
work.

\newsec{Note Added:} While the paper was being typed, a related paper appeared
on the net \ref\daniel{U. Danielsson, A. Guijosa, M. Kruczenski and
B. Sundborg, hep-th/0004187.} which has some overlap with our work.

\listrefs
\end